\documentclass[11pt]{article}
\usepackage{epsfig,subfig,amsmath, amsthm, amscd, amsfonts, amssymb,
graphicx,  color,colortbl,lscape,natbib}
\usepackage[bookmarksnumbered,plainpages,colorlinks,
linkcolor=red, citecolor=blue, citebordercolor={0 0 1}]{hyperref}
\usepackage{multirow}
\usepackage{amsfonts}
\usepackage{amssymb}
\usepackage{graphicx}%
\providecommand{\U}[1]{\protect\rule{.1in}{.1in}}
\providecommand{\U}[1]{\protect\rule{.1in}{.1in}}
\headheight\baselineskip\headsep2\baselineskip\textwidth210mm
\advance\textwidth -2in\textheight287mm\advance\textheight-2in
\theoremstyle{definition}

\let\oldendproof\endproof

\def\endproof{\hfill$\blacksquare$\oldendproof}
\begin{document}

\title{\bf \Large A simulation study of semiparametric estimation in copula models  based on minimum Alpha-Divergence}


\author{ \bf Morteza Mohammadi\footnote{Email: mortezamohammadi408@mail.um.ac.ir}, Mohammad Amini\footnote{Email: m-amini@um.ac.ir (corresponding author)}, and Mahdi Emadi\footnote{Email: emadi@um.ac.ir}  \\ \\
 $^{\ast,\dagger,\ddagger}${Department of Statistics, Faculty of Mathematical Sciences,} 
 \\ {Ferdowsi University of Mashhad, P.O. Box 1159, Mashhad 91775, Iran}
}

\maketitle

\begin{abstract}
The purpose of this paper is to introduce two semiparametric methods for the estimation of copula parameter. These methods are based on minimum Alpha-Divergence between a non-parametric estimation of copula density using local likelihood probit transformation method and a true copula density function. A Monte Carlo study is performed to measure the performance of these methods based on Hellinger distance and Neyman divergence as special cases of Alpha-Divergence. Simulation results are compared to the Maximum Pseudo-Likelihood (MPL) estimation as a conventional estimation method in well-known bivariate copula models. These results show that the proposed method based on Minimum Pseudo Hellinger Distance estimation has a good performance in small sample size and weak dependency situations. The parameter estimation methods are applied to a real data set in Hydrology.
\end{abstract}


\section{Introduction}
The copulas describe the dependence between random vector components.
Unlike marginal and joint distributions that are clearly observable, the copula of a random vector is a hidden dependence 
structure that connects the joint distribution with its margins. 
The copula parameter captures the inherent dependence between the marginal variables and it can be estimated by either parametric or semiparametric methods. 
Maximum likelihood estimation (MLE), which is used to estimate the parameter of any type of model, is the most effective method.
It can also be applied to copula, but the problem becomes complicated as the number of parameters and dimension of copula increases, because the parameters of the margins and copula are estimated simultaneously.
Therefore, MLE is highly affected by misspecification of marginal distributions.

 A rather straightforward way at the cost of lack of efficiency is inference functions for margins (IFM), which is put forward by 
 \cite{Joe.2005}.
 Similar to MLE in this method the margins of the copula are important, 
 because the parameter estimation is dependent on the choice of the marginal distributions.
In IFM method, the parameters are estimated in two stages. 
In the first stage, the parameters of margins are estimated and then the parameters of copula will be evaluated given the values from the first step.
\cite{Genest.et.al.1995} introduce a semiparametric method, known as maximum pseudo-likelihood (MPL) estimation, similar to MLE.
The only difference between this method and MLE is that the data must be converted to pseudo observations. 
The consistency and asymptotic normality of this method is established in their paper. 
They established that this method is efficient for independent copula. 
The results of an extensive simulation studied by \cite{kim.et.al.2007} show that the ML and IFM methods
are non-robust against misspecification of the marginal distributions, 
and that the MPL estimation method performs better than the ML and IFM methods, overall.

The minimum distance (MD) method attains one of the most attractive alternatives to the MLE because the non-parametric estimator of MD has nice robustness properties.
In the case of data containing severe outliers which makes the likelihood-based inference infeasible, the MD method has more appeals. 
 Asymptotic  distributions  of  particular  minimum  distance  estimates were derived by \cite{Millar.1981} for the Cramer-von Mises (CvM) distance; by \cite{Rao.et.al.1975} for the Kolmogorov-Smirnov (KS) distance; by \cite{Beran.1977} for the Hellinger distance.  
\cite{Beran.1977} show that by using minimum Hellinger distance estimators one could obtain robustness properties together with the first-order efficiency.  
%
%
%

 The MD method for copulas has attracted only a little attention in contrast to the MPL and IFM methods.
This paper is closely related to the works of \cite{Tsukahara.2005} and \cite{weib.2011}. 
\cite{Tsukahara.2005} explores the empirical asymptotic behaviour of  CvM and KS distances between the hypothesised and empirical copula in a simulation study. 
He finds that the MPL estimator should be preferred to the MD estimator. 
His analysis is only based on a sample size of 100 and does not include the Gaussian and Student's t (T) copula which are of particular interest in Finance and Hydrology. 
%
%
%
\cite{weib.2011} presented a comprehensive Monte Carlo simulation study on the performance of minimum-distance and maximum-likelihood estimators for bivariate parametric copulas. 
In particular, he considered CvM, KS and $ L^1 $-variants of the CvM-statistic based on the empirical copula process, Kendall's dependence function and Rosenblatt's probability integral transform.

\cite{Tsukahara.2005} proposed the Hellinger distance based on copula density to improve the performance of the MD estimator, 
but did not proceed with it, because it required the estimation of the copula density function.
Hellinger distance is a special case of Alpha-Divergence.
The authors present semiparametric methods based on minimum Alpha-Divergence estimation
 between non-parametric estimation of copula density and true copula density
 which it calls "Minimum Pseudo Alpha-Divergence" (MPAD) estimation.
In this method, the copula density is estimated using local likelihood probit transformation ($\mathcal{LLPT}$) method that was recently suggested by \cite{Geenens.et.al.2017}.
 The purpose of this paper is to present a comprehensive simulation study on the
performance of the MPL estimator and special cases of the MPAD estimator for bivariate parametric copulas.

In what follows, discussions will be restricted to bivariate observations only for simplicity.
The rest of the paper is arranged as follows. 
In Section 2, the preliminaries for copulas and MPL method are described.
The estimation of the copula density function using local likelihood probit transformation method is provided in Section 3.
 In Section 4, the copula parameter estimation based on minimum Alpha-Divergence is introduced.
The simulation results are provided to compare the MPL and MPHD methods in Section 5.
In Section 6, the performance of the considered methods for real data in Hydrology is presented.
Concluding remarks are given in Section 7.

\vspace*{0.1cm}
\section{Preliminaries }
 Some definitions related to a copula function will be briefly reviewed. 
\cite{Sklar.1959} was the primary to display the fundamental concept of the copula.
Let $(X, Y)$ be a continuous random variable with joint cumulative distribution function (cdf) $ F $, then copula $C$ corresponding to $ F $ defined as:
\begin{align}\label{sklar1959}
F(x,y)=C(F_X(x),F_Y(y)), \qquad (x, y) \in R^2, 
\end{align}
where $ F_X $ and $ F_Y $ are the marginal distributions of $ X $ and $ Y$, respectively.
A bivariate copula function $C$ is a cumulative distribution function of random vector $(U, V)$, 
defined on the unit square $  [0,1]^2 $, with uniform marginal distributions as $ U=F_X(X) $ and $ V=F_Y(Y) $.

The authors shall write $ C(u,v;\theta) $ for a family of copulas indexed by the parameter $ \theta $.
If $C(u,v;\theta)$ is an absolutely continuous copula distribution on $ [0,1]^2 $, then its density function is
$c(u,v;\theta)=\frac{\partial^2 C(u,v;\theta)}{\partial u \partial v}$. 
As a result, the relationship between the copula density function ($ c $) and the joint density function ($ f $) of $(X,Y)$ according to equation \eqref{sklar1959} can be represented as
\begin{align}\label{copuladensity2}
 f(x,y)=c(F_X(x),F_Y(y);\theta) f_X(x) f_Y(y), \qquad (x, y) \in R^2,
\end{align}
where $ f_X $ and $ f_Y $ are the marginal density functions of $ X $ and $ Y $, respectively.

Table \ref{table 1} presents summary information of some well-known bivariate copulas such as the parameter space and
 Kendall's tau ($ \tau $) of them.
In this table, Clayton, Gumbel, and Frank copulas belong to the class of Archimedean copulas and Gaussian and T copulas belong to the class of Elliptical copulas.
The copula-based Kendall's tau association for continuous variables $X$ and $ Y$  with copula $C$ is given by
 $ \tau=4\int_{[0,1]^2} C(u,v) dC(u,v)-1 $.

\begin{table}[t]%
\begin{center}
\caption{Some well-known bivariate copulas}\label{table 1}
\centering
\begin{tabular}{ c c c c}
\hline 
\text{Copula} & $ C(u,v;\theta) $ & \text{Parameter Space}&  Kendall's tau \\
\hline 
$Clayton$ & $(u^{-\theta}+v^{-\theta}-1)^{-1/\theta} $ &$\theta\in(-1,+\infty)-\{0\} $ &$ \frac{\theta}{\theta+2} $ \\ 
$Gumbel$  &$\exp\Big\{-\Big[(-\ln u)^\theta+(-\ln v)^\theta\Big]^{1/\theta}\Big\} $ &$\theta\in [1,+\infty)$&$ \frac{\theta-1}{\theta} $  \\
$Frank$  \footnotemark[1] & $ \frac{-1}{\theta} log\Big\{ 1+\frac{(e^{-u\theta}-1)(e^{-v\theta}-1)}{e^{-\theta}-1} \Big\} $ &$\theta\in(-\infty,+\infty) - \{0\} $ &$ 1+\frac{4}{\theta}(D_1(\theta)-1) $ \\ 
$Gaussian$   \footnotemark[2]&$\Phi_2 (\Phi^{-1}(u),\Phi^{-1}(v);\theta)$ &$\theta\in[-1,+1] $ &$ \frac{2}{\pi} arcsin(\theta)$ \\
$T$  \footnotemark[3]& $t_{2,\nu}(t^{-1}_{\nu}(u),t^{-1}_{\nu}(v);\theta) $ &$\theta\in[-1,+1] , \nu>1  $ &$ \frac{2}{\pi} arcsin(\theta)$ \\ 
\hline 
\end{tabular}
\end{center}
\end{table}
\footnotetext[1]{$D_k(\theta)=\frac{k}{\theta^k}\int_{0}^{\theta}\frac{t^k}{e^t-1}dt$.}
\footnotetext[2]{$ \Phi^{-1} $ is the inverse of the standardized univariate Gaussian distribution and $ \Phi_{2} $ is the standardized bivariate Gaussian distribution with correlation parameter  $ \theta $.}
\footnotetext[3]{$ t^{-1}_{\nu} $ is the inverse of the standardized univariate Student's t distribution with $ \nu $ degree of freedom and $ t_{2,\nu} $ is the standardized bivariate Student's t distribution with correlation coefficient $ \theta $ and $\nu$ degree of freedom.}

Let $(X_1, Y_1), (X_2, Y_2), . . . , (X_n, Y_n)$ be a random sample of size $n$ from a pair $(X, Y)$. 
 Empirical copula that was initially introduced by \cite{Deheuvels.1979} defined as
 \begin{align}\label{empriCop}
 C_n(u,v) = \frac{1}{n} \sum_{i=1}^{n} I\{\tilde{U}_i \leq u , \tilde{V}_i \leq v\},
 \end{align}
 where $ \tilde{U}_i= n\hat{F}_X(x_i)/ (n+1) $, $\tilde{V}_i= n\hat{F}_Y(y_i)/ (n+1) $ for $i =1,\cdots,n$, are the pseudo observations and $ \hat{F}_X $ and $ \hat{F}_Y $ are the empirical cumulative distribution function of the observation $ X_i  $ and $ Y_i $, respectively.

\subsection{Semiparametric maximum likelihood estimation}
In view of \eqref{copuladensity2}, the log-likelihood function takes the form
\begin{eqnarray*}
\mathcal{L}(\theta) =\sum_{i=1}^{n} log\Big(c(F(x),G(y);\theta)\Big)+\sum_{i=1}^{n} log\Big( f(x)\Big)+ \sum_{i=1}^{n} log\Big( g(y)\Big).
\end{eqnarray*}
Hence the MLE of $ \theta $, which we denote by $ \hat{\theta}_{ML} $
is the global maximizer of $ \mathcal{L}(\theta) $ and $ \sqrt{n} (\hat{\theta}_{ML}- \theta)  $ converges to a Gaussian distribution with mean zero, where $ \theta $ is the true value.
 Since we assume that the model is correctly specified and hence $ \mathcal{L}(\theta) $ is the correct log-likelihood, it follows that the MLE enjoys some optimality properties and hence is the preferred first option. 
 If the model is not correctly specified so that $ \mathcal{L}(\theta) $ is not the correct log-likelihood, then the
maximizer of $ \mathcal{L}(\theta) $ is not the MLE and hence it may lose its preferred status.

 In MPL method, the marginal distributions have unknown functional forms. 
 Estimation of marginal distributions are estimated non parametrically by their sample empirical distributions. 
Then, $ \theta $ is estimated by the maximizer of the pseudo log-likelihood,
\begin{eqnarray}\label{MPLestim}
\hat{\theta}_{MPL} =\arg \max_{\theta} \sum_{i=1}^{n} log\Big(c(\tilde{U}_i,\tilde{V}_i;\theta)\Big),
\end{eqnarray}
where $ (\tilde{U}_i,\tilde{V}_i) , i =1,\cdots, n$, are the pseudo observations.
The authors shall refer to \eqref{MPLestim} as the maximum pseudo likelihood (MPL) estimator of $ \theta $.
\cite{Genest.et.al.1995} and \cite{Tsukahara.2005} showed that $ \hat{\theta}_{MPL}  $ is consistent estimator.
This non-linear optimization problem can easily be solved by Statistical programming language R or Mathematica.

\vspace*{0.1cm}
\section{Local likelihood probit transformation estimation}
Transformation method was introduced to kernel copula density estimation by \cite{Charpentier.et.al.2007}.
 The simple idea is to transform the data so that it is supported on the full $ R^2 $ (instead of the unit cube). 
On this transformed domain, standard kernel techniques can be used to estimate the density. 
An adequate back-transformation then yields an estimate of the copula density. 
The inverse of the standard Gaussian CDF is most commonly used for the transformation since it is known
that kernel estimators tend to do well for Gaussian random variables.

Let $ (U_i, V_i)_{i=1,...,n} $ are independent and identically distributed observations from the bivariate copula C 
and the purpose is to estimate the corresponding copula density function.
Denote $ \Phi $ as the standard Gaussian distribution and $ \phi $ as its first order derivative. Then 
$ (S_i, T_i)=(\Phi^{-1}(U_i),\Phi^{-1}(V_i)) $
is a random vector with Gaussian margins and copula C. 
According to \eqref{copuladensity2}, the corresponding density function can be written as 
$ f(s,t)=c(\Phi(s),\Phi(t)) \phi(s) \phi(t) $.
Thus,
 an estimation of the copula density function can be given by
\begin{equation}\label{GaussianTransformcopula}
\hat{c }_{n}^{ (\mathcal{PT})}(u,v)=\frac{\hat{f}_{n}(\Phi^{-1}(u),\Phi^{-1}(v))}{\phi(\Phi^{-1}(u)) \phi(\Phi^{-1}(v))}, 
\qquad (u, v) \in (0, 1)^2.
\end{equation}

However, as the $ (U_i, V_i) $ are unavailable and one has to use $ (\hat{S}_i, \hat{T}_i)=(\Phi^{-1}(\hat{U}_i),\Phi^{-1}(\hat{V}_i))$ the pseudo-transformed sample, instead.
As a first natural idea, the standard kernel density estimator for $ \hat{f}_{n} $  in \eqref{GaussianTransformcopula} can be considered as follows:
\begin{equation*}
\hat{f}_{n} (s,t)=\frac{1}{n | \textbf{H}_{ST}|^{\frac{1}{2}}} \sum_{i=1}^{n} 
\textbf{K}\Big( \textbf{H}_{ST}^{-\frac{1}{2}} \Big( 
\begin{matrix}
s-\hat{S}_{i}\\
t-\hat{T}_{i}
\end{matrix}
\Big) \Big),
\end{equation*}
where $ \textbf{K}:R^2\rightarrow R $ is a kernel function, and
 $  \textbf{H}_{ST} = \begin{bmatrix}  b_n & 0 \\ 0 & b_n  \end{bmatrix}$ 
 is a bandwidth matrix. 

This kernel estimator has asymptotic problems at the edges of the distribution support. 
To remedy this problem, local likelihood probit transformation ($\mathcal{LLPT}$) method was recently suggested by \cite{Geenens.et.al.2017}.
Instead of applying the standard kernel estimator, they locally fit a polynomial to the log-density of the transformed sample. 
The advantages of estimating $ f(s,t) $ by local likelihood methods instead of raw kernel density estimation are related to the detailed discussion in  \cite{Geenens.2014}.
This method can fix the boundary issues in a natural way and able to cope with unbounded copula densities.
The notations are similar to ones used in \cite{Geenens.et.al.2017}. 
Recently, \cite{Nagler.2018} with a comprehensive simulation study has shown that $\mathcal{LLPT}$ method for copula density estimation yields very good.

Around $(s,t)\in R^2$ and $ (s^\prime,t^\prime) $ close to $ (s,t) $, the local log-quadratic likelihood estimation of $\log f(s,t)$ from the pseudo-transformed sample is defined as:
\begin{align*}
log f(s^\prime,t^\prime)&= a_{2,0}(s,t) + a_{2,1}(s,t) (s^\prime -s)+a_{2,2}(s,t) (t^\prime -t) \\
& +a_{2,3}(s,t) (s^\prime -s)^2+a_{2,4}(s,t) (t^\prime -t)^2+a_{2,5}(s,t) (s^\prime -s)(t^\prime -t)\\
&\equiv P_{a_2}(s^\prime -s,t^\prime -t).
\end{align*}
The vector $a_2(s,t)\equiv (a_{2,0}(s,t),\cdots,a_{2,5}(s,t))$ 
is then estimated by solving a weighted maximum likelihood problem as
\begin{align*} 
\hat{a}_2(s,t)&=arg \ \max_{a_2} \Big\{ 
\sum_{i=1}^{n} 
\textbf{K}\Big( \textbf{H}_{ST}^{-\frac{1}{2}} \Big( 
\begin{matrix}
s-\hat{S}_{i}\\
t-\hat{T}_{i}
\end{matrix}
\Big) \Big) P_{a_2}(\hat{S}_{i} -s,\hat{T}_{i} -t) \\ 
& - n \int_{R^2} \textbf{K}\Big( \textbf{H}_{ST}^{-\frac{1}{2}} \Big( 
\begin{matrix}
s-s^\prime\\
t-t^\prime
\end{matrix}
\Big) \Big) exp\big(P_{a_2}(s^\prime -s, t^\prime -t)\big) ds^\prime dt^\prime
\Big\}.
\end{align*}
Therefore, the estimation of $ f(s,t) $ is $ {\tilde{f}}^{p}(s,t)=\exp\{ \hat{a}_2(s,t) \} $ and thus
$\mathcal{LLPT}$ estimator of a copula density is
\begin{equation}\label{locallikelihoodprobittransformatio}
\hat{c }_{n}^{ (\mathcal{LLPT})}(u,v)=\frac{{\tilde{f}}^{p}(\Phi^{-1}(u),\Phi^{-1}(v))}{\phi(\Phi^{-1}(u)) \phi(\Phi^{-1}(v))}, \qquad (u, v) \in [0, 1]^2.
\end{equation}

When the underlying density is on $[0, 1]^2$, 
the performance of the kernel estimator depends on the choice of the kernel function and the bandwidth (smoothing parameter).
 For bandwidth choice, a practical approach is to consider the minimization of the AMISE on the level of the transformed data.
In this article, the bandwidth choice based on nearest-neighbor method (see \cite{Geenens.et.al.2017}, Section 4). 

\section{Semiparametric Alpha-Divergence estimation}\label{SPADE}
Initially, \cite{Chernoff.1952}  proposed the Alpha-Divergence, which is a generalization of the KL divergence.
For some Alpha-Divergence investigations see, for example,  
\cite{Amari.Nagaoka.2000}, \cite{Cichocki.Amari.2010}, and \cite{Read.Cressie.2012}.
Alpha-Divergence measure can be derived from Csisz\'ar f-divergence if $ f(t) = \frac{t^\alpha - \alpha (t-1)-1}{\alpha (\alpha-1)}, t\geq 0, \alpha\neq 0, 1$.
The Alpha-Divergence ($ \mathcal{AD} $) between two probability density functions $f_1$ and $f_2$ of a continuous random variable can be defined as:
  \begin{align}\label{Alpha-diver}
\mathcal{AD}_{\alpha}(f_1 \parallel f_2)
=\dfrac{1}{\alpha (\alpha -1)} \Big( \int_{[0, 1]^2 } f_{1}^{\alpha}(x) \ f_{2}^{1-\alpha}(x)dx -1\Big), \qquad \alpha \in R\setminus \{0,1 \}.
\end{align}
The AD divergence is non-negative and true equality to zero holds if and only if $ f_1(x)=f_2(x)$.

If $ \alpha\rightarrow 1 $, the Kullback-Leibler divergence (KLD) can be obtained from equation \eqref{Alpha-diver}. 
The Kullback-Leibler (KL) divergence between two densities $ f_1 $ and $ f_2$ that was introduced by \cite{Kullback.and.Leibler.1951} is given by
 \begin{align*}
KL(f_1||f_2)= \int_{R} \log f_1(x) dF_1 (x) - \int_{R} \log f_2(x) dF_1 (x),
\end{align*}
where $ F_1 (x)=\int_{-\infty}^{x} f_1(t) dt$.
Also, two other special cases of Alpha-Divergence are Hellinger distance and Neyman divergence that will be used in practice.
The well-known Hellinger distance (HD) and Neyman (Neyman Chi-square) divergence (ND) can be obtained from equation \eqref{Alpha-diver} for $ \alpha = 0.5 $ and $ \alpha = 2 $, respectively as
\begin{align}
&HD (f_1 \parallel f_2) =\frac{1}{4} \mathcal{AD}_{1/2}(f_1 \parallel f_2) 
= \frac{1}{2}  \int_{R} (\sqrt{f_1(x)} -\sqrt{f_2(x)})^2  \  dx, \nonumber \\
&ND (f_1 \parallel f_2)  = \mathcal{AD}_{2}(f_1 \parallel f_2) 
= \frac{1}{2}  \int_{R} \frac{(f_1(x) -f_2(x))^2}{f_1(x)}   \  dx. \nonumber
\end{align}

It is well known that maximizing the likelihood is equivalent to minimizing the KL divergence.
Let $c(u,v;\theta)$ be true copula density function associated with copula C.
The MPL estimator is equivalent to minimum pseudo KL divergence (MPKLD) between copula density estimation $ \hat{c}(u,v) $ and true copula density $c(u,v;\theta)$ and given by
\begin{align}
 \hat{\theta}_{MPKLD} &=\arg \min_{\theta}  KL(\hat{c}||c) \nonumber
  \\& =\arg \min_{\theta} \int_{[0, 1]^2 } \log \hat{c}(u,v) dC_n(u,v) - \int_{[0, 1]^2 } \log c(u,v;\theta) dC_n(u,v) \nonumber
  \\& =\arg \max_{\theta}  \int_{[0, 1]^2 } \log c(u,v;\theta) dC_n(u,v) \nonumber
  \\& =\arg \max_{\theta} \frac{1}{n} \sum_{i=1}^{n} log\Big(c(\tilde{U}_i,\tilde{V}_i;\theta)\Big) 
     \equiv \hat{\theta}_{MPL}   \label{MADEMPL11}
\end{align}
The factor $ 1/n $ in the equation \eqref{MADEMPL11} does not affect the attained arg max with respect to $\theta$, and the two approaches MPL and MPKLD gives the same result.
The Alpha-Divergence between copula density estimation $ \hat{c}(u,v) $ and true copula density $c(u,v;\theta)$ to obtain MPAD estimation defined as
$  \hat{\theta}_{MPAD} =\arg \min_{\theta}  \mathcal{AD}(\hat{c}||c) $.

The minimum pseudo Hellinger distance (MPHD) is given by
\begin{align}\label{MPHDestim}
 \hat{\theta}_{MPHD} &=\arg \min_{\theta}  HD(\hat{c}||c) 
  =\arg \min_{\theta} \int_{[0, 1]^2 } \hat{c}(u,v) \Big(1-\sqrt{\frac{c(u,v;\theta)}{\hat{c}(u,v) }}\Big)^2 du dv \nonumber
  \\& =\arg \min_{\theta} \int_{[0, 1]^2 } \Big(1-\sqrt{\frac{c(u,v;\theta)}{\hat{c}(u,v) }}\Big)^2 dC_n(u,v) \nonumber
  \\& =\arg \min_{\theta} \frac{1}{n} \sum_{i=1}^{n}  
                   \Big(1-\sqrt{\frac{c(\tilde{U}_i,\tilde{V}_i;\theta)}{\hat{c}(\tilde{U}_i,\tilde{V}_i) }}\Big)^2. 
\end{align}
Similarly, the minimum pseudo Neyman divergence (MPND) defined as
\begin{align}\label{MPNDestim}
 \hat{\theta}_{MPND} &=\arg \min_{\theta}  ND(\hat{c}||c) 
  =\arg \min_{\theta} \int_{[0, 1]^2 } \hat{c}(u,v) \Big(1-\frac{c(u,v;\theta)}{\hat{c}(u,v) }\Big)^2 du dv \nonumber
  \\& =\arg \min_{\theta} \int_{[0, 1]^2 } \Big(1-{\frac{c(u,v;\theta)}{\hat{c}(u,v) }}\Big)^2 dC_n(u,v) \nonumber
  \\& =\arg \min_{\theta} \frac{1}{n} \sum_{i=1}^{n}  
                   \Big(1-{\frac{c(\tilde{U}_i,\tilde{V}_i;\theta)}{\hat{c}(\tilde{U}_i,\tilde{V}_i) }}\Big)^2. 
\end{align}
%
%
%
In practice, instead of $ \hat{c} $ in equations \eqref{MPHDestim} and \eqref{MPNDestim}, the local likelihood probit transformation estimation of copula density ($ \hat{c }_{n}^{ (\mathcal{LLPT})} $) , which obtain from equation \eqref{locallikelihoodprobittransformatio}, will be used.
\cite{Tsukahara.2005} explores the asymptotic properties of minimum distance estimators based on copula.
He followed \cite{Beran.1984} closely in investigating these properties.

\section{Simulation study}
A simulation study was performed to compare the MPL estimator to the MPHD and MPND estimators as special cases of minimum Alpha-Divergence estimator described in the Section \ref{SPADE}.
All computations were performed using \textbf{copula} and \textbf{kdecop} packages in R software.
 The aim of this simulation study is to compare the true parameter $ \theta $ with the parameter estimate $ \hat{\theta} $, under the assumption that the copula's parametric form is correctly selected.
 This aim is accomplished by comparing the Bias, mean square error (MSE) and relative efficiency (rMSE) of the three approaches of copula parameter estimations that given by
\begin{align*}
& Bias (\hat{\theta})\equiv E(\hat{\theta}) - \theta, 
 \\& MSE (\hat{\theta}) \equiv E(\hat{\theta}- \theta)^2,
 \\& rMSE (\hat{\theta}_1, \hat{\theta}_2)\equiv \sqrt{MSE (\hat{\theta}_2) / MSE (\hat{\theta}_2)}.
\end{align*}

The data are generated from three Archimedean copulas such as Clayton, Gumbel, and Frank
and two Elliptical copulas such as Gaussian and T ($ \nu $=2 and $ \nu $=10) copulas with Kendall's tau 0.1, 0.2, 0.4, 0.6, and 0.8 that are presented in Table \ref{table 1}.
These copulas cover different dependence structures. 
Gaussian and Frank copulas exhibit symmetric and weak tail dependence in both lower and upper tails. 
The Clayton copula exhibits strong left tail dependence and the Gumbel copula has strong right tail dependence. 
In T copula with positive dependency and small degrees of freedom ($ \nu<10 $) tail dependency occurs in both lower and upper tails
and as the degree of freedom increases, dependency in the tail areas decreases (see \cite{Demarta.and.McNeil.2005}). 
Moreover, 1000 Monte Carlo samples of sizes $n = 30$, 75, and 150 are generated from each type of copulas and the three estimates are computed: MPL, MPHD, and MPND.


\begin{table}[!th]
\begin{scriptsize}
\begin{center}
\caption{estimated Bias of the estimators for Archimedean copulas}\label{Bias-Archimedean}
\begin{tabular}{c@{\hspace{2mm}}c@{\hspace{1mm}}cccc@{\hspace{2mm}}cccc@{\hspace{2mm}}cccc}
\hline
\multirow{2}{*}{Copula}  & \multirow{2}{*}{$\tau$} & \multirow{2}{*}{} & \multicolumn{3}{c}{$n=30$}                                             & \multirow{2}{*}{} & \multicolumn{3}{c}{$n=75$}                                             & \multirow{2}{*}{} & \multicolumn{3}{c}{$n=150$}                                            \\ \cline{4-6} \cline{8-10} \cline{12-14} 
                         &                         &                   & $\hat{\theta}_{MPL}$ & $ \hat{\theta}_{MPHD}$ & $ \hat{\theta}_{MPND}$ &                   & $\hat{\theta}_{MPL}$ & $ \hat{\theta}_{MPHD}$ & $ \hat{\theta}_{MPND}$ &                   & $\hat{\theta}_{MPL}$ & $ \hat{\theta}_{MPHD}$ & $ \hat{\theta}_{MPND}$ \\ \hline
\multirow{5}{*}{Clayton} & 0.1                     &                   & 0.0140               & -0.0037                & -0.0124                &                   & 0.0095               & -0.0022                & -0.0088                &                   & 0.0011               & -0.0013                & -0.0014                \\ 
                         & 0.2                     &                   & 0.0288               & -0.0180                & -0.0973                &                   & 0.0216               & -0.0146                & -0.0714                &                   & 0.0107               & -0.0129                & -0.0582                \\ 
                         & 0.4                     &                   & 0.0624               & -0.0516                & -0.1825                &                   & 0.0334               & -0.0376                & -0.1306                &                   & 0.0181               & -0.0228                & -0.1133                \\
                         & 0.6                     &                   & 0.0807               & -0.2256                & -0.4554                &                   & 0.0432               & -0.1633                & -0.3761                &                   & 0.0347               & -0.1119                & -0.2790                \\
                         & 0.8                     &                   & 0.1069               & -0.4127                & -0.8107                &                   & 0.0844               & -0.3835                & -0.6848                &                   & 0.0439               & -0.2381                & -0.5727                \\ \hline
\multirow{5}{*}{Gumbel}  & 0.1                     &                   & 0.0362               & 0.0157                 & -0.0359                &                   & 0.0106               & -0.0091                & 0.0217                 &                   & 0.0017               & -0.0062                & -0.0106                \\ 
                         & 0.2                     &                   & 0.0373               & -0.0219                & -0.0329                &                   & 0.0119               & -0.0113                & -0.0248                &                   & 0.0021               & -0.0076                & -0.0213                \\ 
                         & 0.4                     &                   & 0.0460               & -0.0414                & -0.0622                &                   & 0.0124               & -0.0328                & -0.0575                &                   & 0.0028               & -0.0106                & -0.0432                \\ 
                         & 0.6                     &                   & 0.0730               & -0.2323                & -0.2425                &                   & 0.0157               & -0.1512                & -0.1797                &                   & 0.0045               & -0.1357                & -0.1427                \\ 
                         & 0.8                     &                   & 0.1188               & -0.5503                & -0.5853                &                   & 0.0319               & -0.5195                & -0.5455                &                   & 0.0113               & -0.3847                & -0.4163                \\ \hline
\multirow{5}{*}{Frank}   & 0.1                     &                   & 0.0924               & -0.0331                & -0.0502                &                   & 0.0744               & -0.0229                & -0.0371                &                   & 0.0501               & -0.0163                & -0.0198                \\ 
                         & 0.2                     &                   & 0.1222               & -0.1032                & -0.1172                &                   & 0.0911               & -0.0905                & -0.0947                &                   & 0.0685               & -0.0737                & -0.0850                \\ 
                         & 0.4                     &                   & 0.1436               & -0.1247                & -0.1595                &                   & 0.1271               & -0.1060                & -0.1361                &                   & 0.0894               & -0.0918                & -0.1169                \\ 
                         & 0.6                     &                   & 0.1588               & -0.2594                & -0.2994                &                   & 0.1474               & -0.2376                & -0.2635                &                   & 0.1208               & -0.2004                & -0.2127                \\ 
                         & 0.8                     &                   & 0.1822               & -0.3829                & -0.4165                &                   & 0.1658               & -0.2992                & -0.3487                &                   & 0.1401               & -0.2654                & -0.3183                \\ \hline
\end{tabular}
\end{center}
\end{scriptsize}
\end{table}

\begin{table}[!th]
\begin{scriptsize}
\begin{center}
\caption{estimated Bias of the estimators for Elliptical copulas}\label{Bias-Elliptical}
\begin{tabular}{c@{\hspace{2mm}}c@{\hspace{1mm}}cccc@{\hspace{2mm}}cccc@{\hspace{2mm}}cccc}
\hline
\multirow{2}{*}{Copula}         & \multirow{2}{*}{$\tau$} & \multirow{2}{*}{} & \multicolumn{3}{c}{$n=30$}                                                 & \multirow{2}{*}{} & \multicolumn{3}{c}{$n=75$}                                                 & \multirow{2}{*}{} & \multicolumn{3}{c}{$n=150$}                                                \\ \cline{4-6} \cline{8-10} \cline{12-14} 
                                &                         &                   & $\hat{\theta}_{MPL}$ & $   \hat{\theta}_{MPHD}$ & $   \hat{\theta}_{MPND}$ &                   & $\hat{\theta}_{MPL}$ & $   \hat{\theta}_{MPHD}$ & $   \hat{\theta}_{MPND}$ &                   & $\hat{\theta}_{MPL}$ & $   \hat{\theta}_{MPHD}$ & $   \hat{\theta}_{MPND}$ \\ \hline
\multirow{5}{*}{Gaussian}       & 0.1                     &                   & -0.0171              & -0.0093                  & 0.0109                   &                   & 0.0129               & -0.0063                  & 0.0072                   &                   & -0.0069              & -0.0011                  & -0.0023                  \\
                                & 0.2                     &                   & -0.0188              & -0.0146                  & -0.0227                  &                   & -0.0136              & -0.0123                  & -0.0165                  &                   & -0.0081              & -0.0095                  & -0.0126                  \\ 
                                & 0.4                     &                   & -0.0215              & -0.0192                  & -0.0432                  &                   & -0.0183              & -0.0140                  & -0.0375                  &                   & -0.0023              & -0.0116                  & -0.0296                  \\ 
                                & 0.6                     &                   & -0.0164              & -0.0326                  & -0.0366                  &                   & -0.0065              & -0.0302                  & -0.0338                  &                   & -0.0010              & -0.0227                  & -0.0297                  \\ 
                                & 0.8                     &                   & -0.0022              & -0.0111                  & -0.0529                  &                   & -0.0002              & -0.0073                  & -0.0415                  &                   & -0.0002              & -0.0051                  & -0.0337                  \\ \hline
\multirow{5}{*}{$T(\nu=   2)$}  & 0.1                     &                   & 0.0284               & 0.0128                   & 0.0159                   &                   & 0.0110               & -0.0084                  & 0.0127                   &                   & -0.0039              & -0.0026                  & 0.0115                   \\ 
                                & 0.2                     &                   & -0.0230              & -0.0214                  & -0.0541                  &                   & -0.0138              & -0.0170                  & -0.0437                  &                   & -0.0101              & -0.0124                  & -0.0329                  \\
                                & 0.4                     &                   & -0.0158              & -0.0483                  & -0.0901                  &                   & -0.0147              & -0.0223                  & -0.0813                  &                   & -0.0129              & -0.0162                  & -0.0669                  \\ 
                                & 0.6                     &                   & -0.0148              & -0.0516                  & -0.1126                  &                   & -0.0118              & -0.0463                  & -0.0911                  &                   & -0.0088              & -0.0326                  & -0.0761                  \\ 
                                & 0.8                     &                   & -0.0031              & -0.0488                  & -0.0568                  &                   & -0.0024              & -0.0423                  & -0.0534                  &                   & -0.0017              & -0.0188                  & -0.0232                  \\ \hline
\multirow{5}{*}{$T(\nu=   10)$} & 0.1                     &                   & 0.0258               & 0.0015                   & 0.0129                   &                   & 0.0146               & -0.0011                  & 0.0112                   &                   & 0.0038               & -0.0009                  & -0.0076                  \\
                                & 0.2                     &                   & 0.0065               & -0.0042                  & -0.0268                  &                   & 0.0036               & -0.0031                  & -0.0159                  &                   & 0.0005               & -0.0024                  & -0.0125                  \\ 
                                & 0.4                     &                   & 0.0030               & -0.0384                  & -0.0389                  &                   & 0.0011               & -0.0268                  & -0.0313                  &                   & 0.0003               & -0.0124                  & -0.0236                  \\
                                & 0.6                     &                   & -0.0025              & -0.0460                  & -0.0485                  &                   & 0.0009               & -0.0314                  & -0.0375                  &                   & 0.0007               & -0.0194                  & -0.0317                  \\
                                & 0.8                     &                   & -0.0011              & -0.0163                  & -0.0427                  &                   & 0.0002               & -0.0141                  & -0.0206                  &                   & 0.0001               & -0.0095                  & -0.0143                  \\ \hline
\end{tabular}
\end{center}
\end{scriptsize}
\end{table}


\begin{table}[!th]
\begin{scriptsize}
\begin{center}
\caption{estimated MSE of the estimators for Archimedean copulas}\label{MSE-Archimedean}
\begin{tabular}{c@{\hspace{2mm}}c@{\hspace{1mm}}cccc@{\hspace{2mm}}cccc@{\hspace{2mm}}cccc}
\hline
\multirow{2}{*}{Copula}  & \multirow{2}{*}{$\tau$} & \multirow{2}{*}{} & \multicolumn{3}{c}{$n=30$}                                             & \multirow{2}{*}{} & \multicolumn{3}{c}{$n=75$}                                             & \multirow{2}{*}{} & \multicolumn{3}{c}{$n=150$}                                            \\ \cline{4-6} \cline{8-10} \cline{12-14} 
                         &                         &                   & $\hat{\theta}_{MPL}$ & $ \hat{\theta}_{MPHD}$ & $ \hat{\theta}_{MPND}$ &                   & $\hat{\theta}_{MPL}$ & $ \hat{\theta}_{MPHD}$ & $ \hat{\theta}_{MPND}$ &                   & $\hat{\theta}_{MPL}$ & $ \hat{\theta}_{MPHD}$ & $ \hat{\theta}_{MPND}$ \\ \hline
\multirow{5}{*}{Clayton} & 0.1                     &                   & 0.0791               & 0.0396                 & 0.0742                 &                   & 0.0469               & 0.0256                 & 0.0437                 &                   & 0.0161               & 0.0131                 & 0.0181                 \\ 
                         & 0.2                     &                   & 0.0944               & 0.0689                 & 0.0956                 &                   & 0.0533               & 0.0428                 & 0.0632                 &                   & 0.0232               & 0.0216                 & 0.0298                 \\
                         & 0.4                     &                   & 0.1092               & 0.0818                 & 0.1206                 &                   & 0.0736               & 0.0622                 & 0.1004                 &                   & 0.0341               & 0.0525                 & 0.0737                 \\
                         & 0.6                     &                   & 0.2121               & 0.2925                 & 0.3135                 &                   & 0.1391               & 0.2312                 & 0.2402                 &                   & 0.0834               & 0.1753                 & 0.2002                 \\ 
                         & 0.8                     &                   & 0.5243               & 0.8571                 & 0.8686                 &                   & 0.4549               & 0.8129                 & 0.8345                 &                   & 0.3227               & 0.7778                 & 0.7902                 \\ \hline
\multirow{5}{*}{Gumbel}  & 0.1                     &                   & 0.0282               & 0.0164                 & 0.0260                 &                   & 0.0110               & 0.0087                 & 0.0103                 &                   & 0.0055               & 0.0048                 & 0.0082                 \\ 
                         & 0.2                     &                   & 0.0349               & 0.0226                 & 0.0387                 &                   & 0.0199               & 0.0165                 & 0.0236                 &                   & 0.0086               & 0.0079                 & 0.0159                 \\
                         & 0.4                     &                   & 0.0486               & 0.0342                 & 0.0603                 &                   & 0.0285               & 0.0260                 & 0.0370                 &                   & 0.0121               & 0.0216                 & 0.0278                 \\ 
                         & 0.6                     &                   & 0.1077               & 0.1185                 & 0.1453                 &                   & 0.0595               & 0.0863                 & 0.0894                 &                   & 0.0254               & 0.0537                 & 0.0640                 \\ 
                         & 0.8                     &                   & 0.4591               & 0.7942                 & 0.8325                 &                   & 0.3228               & 0.6535                 & 0.6886                 &                   & 0.1488               & 0.3877                 & 0.3988                 \\ \hline
\multirow{5}{*}{Frank}   & 0.1                     &                   & 0.5431               & 0.4164                 & 0.5143                 &                   & 0.4390               & 0.3680                 & 0.4525                 &                   & 0.2375               & 0.2119                 & 0.2596                 \\ 
                         & 0.2                     &                   & 0.5950               & 0.5167                 & 0.5859                 &                   & 0.4520               & 0.4206                 & 0.4767                 &                   & 0.2554               & 0.2611                 & 0.2997                 \\
                         & 0.4                     &                   & 0.6116               & 0.5691                 & 0.6437                 &                   & 0.4775               & 0.4692                 & 0.5319                 &                   & 0.2693               & 0.2918                 & 0.3487                 \\ 
                         & 0.6                     &                   & 0.6642               & 0.6984                 & 0.7158                 &                   & 0.4831               & 0.5742                 & 0.5983                 &                   & 0.3207               & 0.4379                 & 0.5157                 \\ 
                         & 0.8                     &                   & 0.8096               & 0.8749                 & 0.8967                 &                   & 0.6711               & 0.8494                 & 0.8807                 &                   & 0.4098               & 0.7760                 & 0.8616                 \\ \hline
\end{tabular}
\end{center}
\end{scriptsize}
\end{table}

\begin{table}[!th]
\begin{scriptsize}
\begin{center}
\caption{estimated MSE of the estimators for Elliptical copulas}\label{MSE-Elliptical}
\begin{tabular}{c@{\hspace{2mm}}c@{\hspace{1mm}}cccc@{\hspace{2mm}}cccc@{\hspace{2mm}}cccc}
\hline
\multirow{2}{*}{Copula}         & \multirow{2}{*}{$\tau$} & \multirow{2}{*}{} & \multicolumn{3}{c}{$n=30$}                                             & \multirow{2}{*}{} & \multicolumn{3}{c}{$n=75$}                                             & \multirow{2}{*}{} & \multicolumn{3}{c}{$n=150$}                                            \\ \cline{4-6} \cline{8-10} \cline{12-14} 
                                &                         &                   & $\hat{\theta}_{MPL}$ & $ \hat{\theta}_{MPHD}$ & $ \hat{\theta}_{MPND}$ &                   & $\hat{\theta}_{MPL}$ & $ \hat{\theta}_{MPHD}$ & $ \hat{\theta}_{MPND}$ &                   & $\hat{\theta}_{MPL}$ & $ \hat{\theta}_{MPHD}$ & $ \hat{\theta}_{MPND}$ \\ \hline
\multirow{5}{*}{Gaussian}       & 0.1                     &                   & 0.0421               & 0.0218                 & 0.0255                 &                   & 0.0178               & 0.0147                 & 0.0196                 &                   & 0.0075               & 0.0071                 & 0.0112                 \\ 
                                & 0.2                     &                   & 0.0270               & 0.0161                 & 0.0216                 &                   & 0.0141               & 0.0124                 & 0.0158                 &                   & 0.0070               & 0.0068                 & 0.0108                 \\ 
                                & 0.4                     &                   & 0.0220               & 0.0141                 & 0.0189                 &                   & 0.0109               & 0.0098                 & 0.0138                 &                   & 0.0048               & 0.0062                 & 0.0117                 \\ 
                                & 0.6                     &                   & 0.0085               & 0.0101                 & 0.0126                 &                   & 0.0033               & 0.0061                 & 0.0071                 &                   & 0.0015               & 0.0032                 & 0.0048                 \\ 
                                & 0.8                     &                   & 0.0047               & 0.0069                 & 0.0094                 &                   & 0.0020               & 0.0044                 & 0.0053                 &                   & 0.0011               & 0.0027                 & 0.0038                 \\ \hline
\multirow{5}{*}{$T(\nu=   2)$}  & 0.1                     &                   & 0.0442               & 0.0322                 & 0.0343                 &                   & 0.0261               & 0.0211                 & 0.0337                 &                   & 0.0204               & 0.0186                 & 0.0296                 \\ 
                                & 0.2                     &                   & 0.0372               & 0.0305                 & 0.0333                 &                   & 0.0205               & 0.0194                 & 0.0310                 &                   & 0.0122               & 0.0160                 & 0.0266                 \\ 
                                & 0.4                     &                   & 0.0324               & 0.0276                 & 0.0327                 &                   & 0.0163               & 0.0172                 & 0.0280                 &                   & 0.0088               & 0.0142                 & 0.0217                 \\ 
                                & 0.6                     &                   & 0.0173               & 0.0248                 & 0.0279                 &                   & 0.0066               & 0.0105                 & 0.0219                 &                   & 0.0035               & 0.0089                 & 0.0174                 \\ 
                                & 0.8                     &                   & 0.0042               & 0.0084                 & 0.0139                 &                   & 0.0031               & 0.0083                 & 0.0115                 &                   & 0.0013               & 0.0039                 & 0.0082                 \\ \hline
\multirow{5}{*}{$T(\nu=   10)$} & 0.1                     &                   & 0.0292               & 0.0251                 & 0.0282                 &                   & 0.0218               & 0.0199                 & 0.0241                 &                   & 0.0131               & 0.0126                 & 0.0197                 \\ 
                                & 0.2                     &                   & 0.0275               & 0.0245                 & 0.0273                 &                   & 0.0167               & 0.0159                 & 0.0229                 &                   & 0.0091               & 0.0115                 & 0.0159                 \\ 
                                & 0.4                     &                   & 0.0242               & 0.0226                 & 0.0249                 &                   & 0.0139               & 0.0136                 & 0.0204                 &                   & 0.0066               & 0.0090                 & 0.0138                 \\ 
                                & 0.6                     &                   & 0.0096               & 0.0178                 & 0.0182                 &                   & 0.0065               & 0.0141                 & 0.0169                 &                   & 0.0032               & 0.0076                 & 0.0111                 \\ 
                                & 0.8                     &                   & 0.0044               & 0.0091                 & 0.0116                 &                   & 0.0025               & 0.0062                 & 0.0094                 &                   & 0.0011               & 0.0033                 & 0.0063                 \\ \hline
\end{tabular}
\end{center}
\end{scriptsize}
\end{table}


\begin{table}[!th]
\begin{scriptsize}
\begin{center}
\caption{estimated MSE of MPL estimator relative to the MPHD and MPND estimators (rMSE) in percent for Archimedean copulas}\label{rMSE-Archimedean}
\begin{tabular}{cccccccccc}
\hline
\multirow{2}{*}{Copula}  & \multirow{2}{*}{$\tau$} & \multirow{2}{*}{} & \multicolumn{3}{c}{$   rMSE(\hat{\theta}_{MPL}, \hat{\theta}_{MPHD})$} & \multirow{2}{*}{} & \multicolumn{3}{c}{$   rMSE(\hat{\theta}_{MPL}, \hat{\theta}_{MPND})$} \\ \cline{4-6} \cline{8-10} 
                         &                         &                   & $n=30$                & $n=75$                & $n=150$                &                   & $n=30$                & $n=75$                & $n=150$                \\ \hline
\multirow{5}{*}{Clayton} & 0.1                     &                   & 70.8                  & 73.9                  & 90.2                   &                   & 96.9                  & 96.5                  & 106.1                  \\ 
                         & 0.2                     &                   & 85.4                  & 89.6                  & 96.5                   &                   & 100.7                 & 108.9                 & 113.3                  \\ 
                         & 0.4                     &                   & 86.6                  & 91.9                  & 124.1                  &                   & 105.1                 & 116.8                 & 147.0                  \\ 
                         & 0.6                     &                   & 117.4                 & 128.9                 & 145.0                  &                   & 121.6                 & 131.4                 & 154.9                  \\ 
                         & 0.8                     &                   & 127.9                 & 133.7                 & 155.2                  &                   & 128.7                 & 135.4                 & 156.5                  \\ \hline
\multirow{5}{*}{Gumbel}  & 0.1                     &                   & 76.3                  & 89.0                  & 93.7                   &                   & 95.9                  & 96.9                  & 122.7                  \\ 
                         & 0.2                     &                   & 80.5                  & 91.0                  & 95.8                   &                   & 105.3                 & 108.8                 & 135.8                  \\ 
                         & 0.4                     &                   & 84.0                  & 95.5                  & 133.7                  &                   & 111.4                 & 113.8                 & 151.9                  \\ 
                         & 0.6                     &                   & 104.9                 & 120.4                 & 145.3                  &                   & 116.1                 & 122.6                 & 158.6                  \\ 
                         & 0.8                     &                   & 131.5                 & 142.3                 & 161.4                  &                   & 134.7                 & 146.1                 & 163.7                  \\ \hline
\multirow{5}{*}{Frank}   & 0.1                     &                   & 87.6                  & 91.6                  & 94.4                   &                   & 97.3                  & 101.5                 & 104.5                  \\ 
                         & 0.2                     &                   & 93.2                  & 96.5                  & 101.1                  &                   & 99.2                  & 102.7                 & 108.3                  \\ 
                         & 0.4                     &                   & 96.5                  & 99.1                  & 104.1                  &                   & 102.6                 & 105.5                 & 113.8                  \\ 
                         & 0.6                     &                   & 102.5                 & 109.0                 & 116.9                  &                   & 103.8                 & 111.3                 & 126.8                  \\ 
                         & 0.8                     &                   & 104.0                 & 112.5                 & 137.6                  &                   & 105.2                 & 114.6                 & 145.0                  \\ \hline
\end{tabular}
\end{center}
\end{scriptsize}
\end{table}

\begin{table}[!th]
\begin{scriptsize}
\begin{center}
\caption{estimated MSE of MPL estimator relative to the MPHD and MPND estimators (rMSE) in percent for Elliptical copulas}\label{rMSE-Elliptical}
\begin{tabular}{cccccccccc}
\hline
\multirow{2}{*}{Copula}       & \multirow{2}{*}{$\tau$} & \multirow{2}{*}{} & \multicolumn{3}{c}{$   rMSE(\hat{\theta}_{MPL}, \hat{\theta}_{MPHD})$} & \multirow{2}{*}{} & \multicolumn{3}{c}{$   rMSE(\hat{\theta}_{MPL}, \hat{\theta}_{MPND})$} \\ \cline{4-6} \cline{8-10} 
                              &                         &                   & $n=30$                & $n=75$                & $n=150$                &                   & $n=30$                & $n=75$                & $n=150$                \\ \hline
\multirow{5}{*}{Gaussian}     & 0.1                     &                   & 72.0                  & 90.8                  & 97.4                   &                   & 77.8                  & 104.8                 & 122.1                  \\ 
                              & 0.2                     &                   & 77.2                  & 93.8                  & 99.1                   &                   & 89.5                  & 105.8                 & 124.7                  \\ 
                              & 0.4                     &                   & 80.3                  & 95.1                  & 113.2                  &                   & 92.9                  & 112.6                 & 155.5                  \\ 
                              & 0.6                     &                   & 109.1                 & 136.0                 & 146.9                  &                   & 121.4                 & 147.1                 & 178.8                  \\ 
                              & 0.8                     &                   & 120.7                 & 148.9                 & 153.8                  &                   & 140.8                 & 164.6                 & 182.9                  \\ \hline
\multirow{5}{*}{$T(\nu= 2)$}  & 0.1                     &                   & 85.4                  & 90.0                  & 95.4                   &                   & 88.1                  & 113.5                 & 120.5                  \\ 
                              & 0.2                     &                   & 90.6                  & 97.3                  & 114.3                  &                   & 94.6                  & 123.1                 & 147.3                  \\ 
                              & 0.4                     &                   & 92.3                  & 102.7                 & 127.1                  &                   & 100.5                 & 131.0                 & 157.2                  \\ 
                              & 0.6                     &                   & 119.9                 & 126.0                 & 159.5                  &                   & 127.2                 & 182.0                 & 222.8                  \\ 
                              & 0.8                     &                   & 141.0                 & 163.5                 & 172.0                  &                   & 181.6                 & 192.1                 & 250.2                  \\ \hline
\multirow{5}{*}{$T(\nu= 10)$} & 0.1                     &                   & 92.7                  & 95.5                  & 98.1                   &                   & 98.2                  & 105.0                 & 122.5                  \\ 
                              & 0.2                     &                   & 94.5                  & 97.7                  & 112.5                  &                   & 99.7                  & 117.2                 & 132.3                  \\ 
                              & 0.4                     &                   & 96.6                  & 99.0                  & 117.2                  &                   & 101.4                 & 121.0                 & 145.3                  \\ 
                              & 0.6                     &                   & 136.2                 & 147.5                 & 154.4                  &                   & 137.4                 & 161.1                 & 185.9                  \\ 
                              & 0.8                     &                   & 144.3                 & 157.1                 & 169.1                  &                   & 162.9                 & 193.2                 & 234.2                  \\ \hline
\end{tabular}
\end{center}
\end{scriptsize}
\end{table}

\subsection{Results}
Results of the simulation study are presented in Tables \ref{Bias-Archimedean}-\ref{rMSE-Elliptical}.
These tables present the Bias and MSE relative to the three estimators of the respective copulas for different values of sample sizes and Kendall's tau.
The simulation procedure was performed for the positive and negative values of Kendall's tau and according to the symmetry of the obtained results, the results have been reported only for positive values of Kendall's tau.
As the results for the sample sizes greater than 150 were in line with our expectation that the increase in sample size will improve the parameter estimation, the corresponding results were omitted from the tables for brevity. 
Also, the results show that the MPL method outperforms MPHD and MPND for sample sizes greater than 150.
 The results for the T copula with 4 and 7 degrees of freedom were omitted as well as the results did not differ from those for the two other T copulas with 2 and 10 degrees of freedom.

The results given in Tables \ref{Bias-Archimedean}-\ref{rMSE-Elliptical} show that
estimated Bias and MSE of parameter estimation of the Archimedean and Elliptical copulas decrease as sample size increases and
 parameter estimates improve.
 The estimated Bias and MSE of parameter estimation increase with increasing Kendall's tau for Archimedean copulas.
Also, estimated MSE of parameter estimation decrease with increasing Kendall's tau, whereas  
 estimated Bias of parameter estimation has no clear trend for Elliptical copulas.
 Furthermore, the results for estimated MSE of MPL estimator relative to the MPHD and MPND estimators (rMSE) in percent for Archimedean and Elliptical copulas in Tables \ref{rMSE-Archimedean}-\ref{rMSE-Elliptical} show that rMSE increase with increasing sample size or Kendall's tau.

The results given in Tables \ref{Bias-Archimedean}-\ref{MSE-Elliptical} show that the MPL yields the best results for the large sample size ($ n\geq 100 $) and high dependency ($ \tau\geq 0.5 $). 
For the small sample size ($ n < 100 $) and weak dependency ($ \tau<0.5 $) , Minimum Hellinger distance estimation outperforms MPL estimation method. 
Among the two new minimum distance estimators, the results show that $ \hat{\theta}_{MPHD}$  is better than $ \hat{\theta}_{MPND}$ based on MSE in always.
This advantage for $ \hat{\theta}_{MPHD}$ is clearer in Archimedean copulas than in Elliptical copulas.
 Thus, there is no evident reason why one would be inclined to use an  $ \hat{\theta}_{MPND}$.
In addition to these results, the estimated bias seem to be considerably higher for Archimedean copulas than for Elliptical copulas. 
  In all tables, the biases of the MPL estimators are almost always lower than the biases of the MPHD and MPND estimators for the large sample size ($n>100$).
Finally, it is necessary to note that although the time required to compute the MPHD method is longer than the MPL method, the MPHD method has accurate and acceptable results for small sample size and weak dependency.

\section{Application in Hydrology}
An application of estimation methods is demonstrated to a given dataset in Hydrology.
\cite{Wong.et.al.2008} established a joint distribution function of drought intensity, duration, 
and severity by using Gaussian and Gumbel copulas. 
 \cite{Song.and.Singh.2010a} used several meta-elliptical copulas in drought analysis and found that meta-Gaussian and T copula had a better fit. 
\cite{Ma.et.al.2013} investigated the drought events in the Weihe river basin and selected the Gaussian and T copulas to model the joint distribution among drought duration, severity, and peaks. 
Recently, a very comprehensive book on the application of copula in Hydrology has been published 
by \cite{Chen.and.Guo.2019} and the concepts in this section are taken from this book.

\cite{McKee.et.al.1993} proposed the concept of standardized precipitation index (SPI) based on the long-term precipitation record
for a specific period such as 1, 3, 6, 12, months, etc.
\cite{Guttman.1998} recommended the use of SPI as a primary drought index because it is simple, spatially invariant in its interpretation, and probabilistic. 
Therefore, the SPI series is used for this article.
Fitting this long-term precipitation record to a probability distribution is the first step to calculate SPI series. 
Once the probability distribution is determined, the cumulative probability of observed precipitation is computed and then inverse transformed by a standard Gaussian distribution is equal to SPI series.
A drought event is thus defined as a continuous period in which the SPI is below 0.

The objective of this section is the estimation of copula parameter between drought characteristics (events) based on SPI, including 
drought duration, drought severity, and drought interval time.
Drought characteristics are recognized as important factors in water resource planning and management.
Drought duration ($D_d$) is defined as the number of consecutive intervals (months)
where SPI remains below the threshold value 0 (see \cite{Shiau.2006}).
Drought severity ($S_d$) is defined as a cumulative SPI value during a drought period, $S_d=\sum_{i=1}^{D_d} SPI_i$ where $SPI_i$ means the SPI value in the ith month (see \cite{Mishra.and.Singh.2010}).
The drought interval time ($I_d$) is defined as the period elapsing from the initiation of drought to 
the beginning of the next drought (see \cite{Song.and.Singh.2010b}).

\begin{figure}[!h]
\begin{center}
\includegraphics[width=\linewidth]{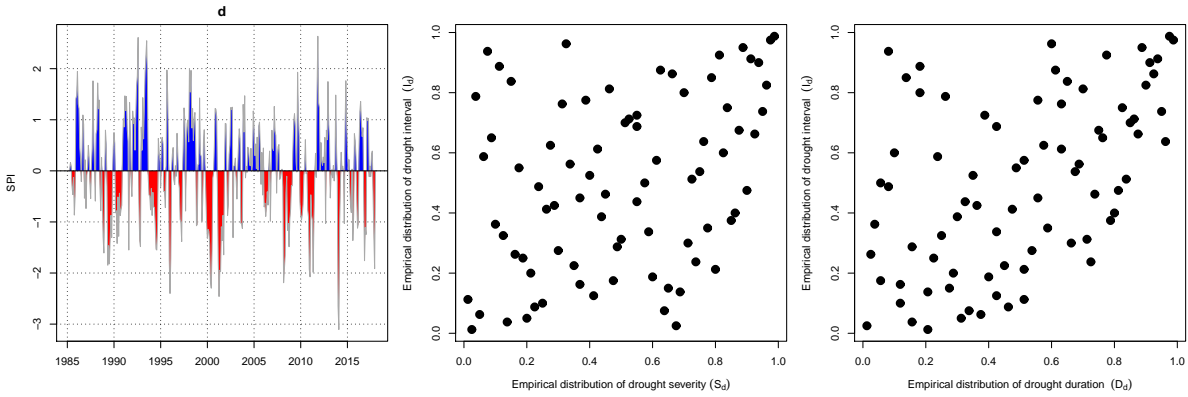}
\caption{The 1-month SPI time series for the Masshad station [left panel] and
 scatter plots for the empirical distributions of pair $  (S_d,I_d)  $ [middle panel] and pair $  (D_d,I_d)  $ [right panel]} \label{SPIMashhad}
\end{center}
\end{figure}

The monthly precipitation data of Mashhad station, located in Iran, from 1985 to 2017 
(http://www.irimo.ir/eng/index.php) 
is used as an example to illustrate the proposed methodology. 
The monthly precipitation of Mashhad can be fitted by a gamma distribution. 
The monthly SPI series is then calculated and demonstrated in Figure \ref{SPIMashhad} (left panel) for this 33-year period.
Thereupon,  the drought variables with sample size 79 are obtained.
 The pseudo observations of $S_d$, $D_d$, and $I_d$ are used to copula parameter estimation.
The estimation of sample version of  Kendall's tau correlation coefficient ($ \hat{\tau}_n $) of drought variables is calculated.
The results confirm that two pairs $  (S_d, I_d)  $ and $  (D_d, I_d)  $ have positive and weak dependency.
 The values ($ \hat{\tau}_n $) for two pairs $  (S_d, I_d)  $ and $  (D_d, I_d)  $ of drought variables are given in Table \ref{SPI-table-Mashhad}.

A goodness of fit testing procedure based on parameter estimations methods is applied.
In the large scale Monte Carlo experiments carried out by \cite{Genest.et.al.2009}, the CvM statistic as
\begin{align*}
 S_n =n \int_{[0,1]^2} \Big(C_n (u,v)- C_{\hat{\theta}}(u,v)\Big)^2 dC_n (u,v) = \sum_{i=1}^{n} \Big(C_n (\tilde{U}_i,\tilde{V}_i)- C_{\hat{\theta}}(\tilde{U}_i,\tilde{V}_i)\Big)^2,
\end{align*}
gave the best results overall, where $C_n$ is the empirical copula defined in \eqref{empriCop} and $ C_{\hat{\theta}} $ is an estimator of C under the hypothesis that $H_0 : C \in {C_{\theta}} $ holds. 
The estimators $ \hat{\theta} $ of $ \theta $ appearing  in \eqref{MPLestim} and \eqref{MPHDestim}.
An approximate P-Value for $  S_n $ can be obtained by means of a parametric bootstrap-based procedure as described in \cite{Genest.et.al.2009}

One of the challenges that we face is the specification of a suitable copula. 
Since there are a large number of copulas, specifying one that would suit a particular case in practice is not easy. 
Therefore, a reasonable strategy is to consider different copulas and evaluate their goodness of fits. 
To this end, the Archimedean and Elliptical copulas in Table \ref{table 1} are considered  
that have attracted considerable interest because of its flexibility and simplicity. 
The diagnostic checks to investigate the dependence structure for pairs $  (S_d, I_d) $ and $  (D_d, I_d)  $ suggested that Gumbel and Gaussian copulas fit well and better than the others considered.
The Gumbel and Gaussian copulas are fitted  by the MPL and MPHD methods.
 The estimates and various relevant quantities are presented in Table \ref{SPI-table-Mashhad}.

\begin{table}[]
\begin{scriptsize}
\begin{center}
\caption{Parameter estimates and summary statistics for the SPI-Mashhad data}\label{SPI-table-Mashhad}
\centering
\begin{tabular}{cccccccc}
\hline
Pair                         & Copula                  & Method & $\hat{\theta}$ & $\tau(\hat{\theta})$ & $S_n$ & P-Value & AIC\\ \hline
                             & Gumbel & MPL    & 1.4176         & 0.2946                         & 0.0234 & 0.6287&
-16.1803
  \\ 
$(S_d, I_d)$                    &                         & MPHD   & 1.3047         & 0.2335                         & 0.0212 & 0.6418 &
-17.0441
 \\ \cline{2-8} 
($\hat{\tau}_n=0.2394$) & Gaussian & MPL    & 0.4312         & 0.2838                         & 0.0332 & 0.4032&
-11.2319
  \\ 
                             &                         & MPHD   & 0.3694         & 0.2409                         & 0.0311 & 0.4203 &
-11.9615
 \\ \hline

                             & Gumbel & MPL    & 1.5940         & 0.3726                         & 0.0369 & 0.3165 & 
-27.0587
  \\  
$(D_d, I_d)$                    &                         & MPHD   & 1.5608         & 0.3593                         & 0.0336 & 0.3390 & 
-27.4128
  \\ \cline{2-8} 
($\hat{\tau}_n=0.3634$) & Gaussian & MPL    & 0.5535         & 0.3735                         & 0.0392 & 0.2308 & 
-23.1681
  \\  
                             &                         & MPHD   & 0.5303         & 0.3558                         & 0.0375 & 0.2639 & 
-23.4688
  \\ \hline
\end{tabular}
\end{center}
\end{scriptsize}
\end{table}

The scatter plots for the empirical distributions of pair $  (S_d, I_d)  $ [middle panel] and pair $  (D_d, I_d)  $ [right panel] are shown  in Figure \ref{SPIMashhad}.
This figure shows that the points tend to concentrate near (1, 1). 
Thus, the Gumbel copula that have upper tail dependence appears to be more appropriate for both two pairs.
On the other hand, according to the values ​​of the Akaike Information Criterion (AIC) in Table \ref{SPI-table-Mashhad}, it can be concluded that for both pairs $  (S_d, I_d) $ and $  (D_d, I_d)  $, the Gumbel copula is better suitable than Gaussian copula, because it has the least value of AIC.
The P-Values and values of statistic $ S_n $ can be used to compare the goodness of fits.
These are given here just as a point of reference but we recognize that they do not have the usual meaning of the P-Value.
The large P-Values, for pair $  (S_d, I_d) $ based on $  S_n $ would be 0.6418 for the Gumbel copula with parameter estimation by MPHD.
Also,  the large P-Values, for pair $  (D_d, I_d) $ based on $  S_n $ would be 0.3390 for the Gumbel copula with parameter estimation by MPHD.
The values of the copula parameter are difficult to interpret, but the corresponding values of the Kendall's tau have more intuitive interpretations.
By using the relations in Table \ref{table 1}, the values the Kendall's tau corresponding to the different estimates of $ \theta $ ($\tau(\hat{\theta})$) are given in Table \ref{SPI-table-Mashhad}. 
 Note that for pair $  (S_d, I_d) $, the Gumbel copula based on MPHD method has $ \hat{\theta}_{MPHD}=1.3047 $ and $\tau(\hat{\theta})=0.2335 $.
The fact that $\tau(\hat{\theta})$ is nearly identical to the non-parametric sample estimate, $\hat{\tau}_n=0.2394$, implies that the MPHD approach handles this dependency aspect well.
This provides additional support to previous observation that the MPHD method estimated well and better than the MPL.
Overall, the results suggest that the Gumbel copula estimated by MPHD provides an acceptable fit for both pairs of drought variables.

\section{Conclusion}
In this paper, two methods of copula parameter estimation based on Alpha-Divergence were presented for some bivariate Archimedean and Elliptical copulas.
The minimum of Kullback-Leibler divergence, Hellinger distance, and Neyman Divergence as special cases of Alpha-Divergence based on pseudo observations were used to obtain the copula parameter estimation.
The simulation results suggests that the minimum pseudo Hellinger distance estimation method has good performance in 
small sample size ($ n < 100 $) and weak dependency ($ \tau<0.5 $) situations when compared with the MPL estimation methods for Archimedean and Elliptical copulas.
Also, the simulation results show that $ \hat{\theta}_{MPHD}$  is better than $ \hat{\theta}_{MPND}$ in almost always.
The estimation methods were developed in the Goodness of fit test based on CvM distance for a data set in Hydrology and the results show that the MPHD method is more accurate than MPL method.



\begin{thebibliography}{3}


%
%
%


\bibitem[Amari and Nagaoka (2000)]{Amari.Nagaoka.2000}
Amari, S. I., and Nagaoka, H. (2000). \textit{Methods of information geometry} (Vol. 191). American Mathematical Society.

\bibitem[Beran (1977)]{Beran.1977}
Beran, R. (1977). \textit{Minimum Hellinger distance estimates for parametric models}. The Annals of Statistics, \textbf{5}(3), 445-463.


\bibitem[Beran (1984)]{Beran.1984}
Beran, R. (1984). \textit{30 Minimum distance procedures}. Handbook of statistics, \textbf{4}, 741-754.

\bibitem[Cichocki and Amari (2010)]{Cichocki.Amari.2010}
Cichocki, A., and Amari, S. I. (2010). \textit{Families of alpha-beta-and gamma-divergences: Flexible and robust measures of similarities}. Entropy, \textbf{12}(6), 1532-1568.

\bibitem[Charpentier et al. (2007)]{Charpentier.et.al.2007}
Charpentier, A., Fermanian, J. D., and Scaillet, O. (2007). The estimation of copulas: Theory and practice. \textit{Copulas: From theory to application in finance}, 35-64.


\bibitem[Chen and Guo (2019)]{Chen.and.Guo.2019}
Chen, L., and Guo, S. (2019). \textit{Copulas and Its Application in Hydrology and Water Resources}. Singapore: Springer.



\bibitem[Chernoff (1952)]{Chernoff.1952}
Chernoff, H. (1952). \textit{A measure of asymptotic efficiency for tests of a hypothesis based on the sum of observations}. The Annals of Mathematical Statistics, \textbf{23}(4), 493-507.




\bibitem[Deheuvels (1979)]{Deheuvels.1979}
Deheuvels, P. (1979). \textit{La fonction de dependence empirique et ses proprietes, Un test non parametrique d'independance}. Bulletin de la classe des sciences, Academie Royale de Belgique, \textbf{5}(65), 274-292.

\bibitem[Demarta and McNeil (2005)]{Demarta.and.McNeil.2005}
Demarta, S., and McNeil, A. J. (2005). \textit{The t copula and related copulas}. International statistical review, \textbf{73}(1), 111-129.


%



\bibitem[Geenens (2014)]{Geenens.2014}
Geenens, G. (2014). \textit{Probit transformation for kernel density estimation on the unit interval}. Journal of the American Statistical Association, \textbf{109}(505), 346-358.


\bibitem[Geenens et al. (2017)]{Geenens.et.al.2017}
Geenens, G., Charpentier, A., and Paindaveine, D. (2017). \textit{Probit transformation for nonparametric kernel estimation of the copula density}. Bernoulli, \textbf{23}(3), 1848-1873.


\bibitem[Genest et al. (1995)]{Genest.et.al.1995}
Genest, C., Ghoudi, K., and Rivest, L. P. (1995). \textit{A semiparametric estimation procedure of dependence parameters in multivariate families of distributions}. Biometrika, \textbf{82}(3), 543-552.


%
%
%


\bibitem[Genest et al. (2009)]{Genest.et.al.2009}
Genest, C., Rémillard, B., and Beaudoin, D. (2009). \textit{Goodness-of-fit tests for copulas: A review and a power study}. Insurance: Mathematics and Economics, \textbf{44}(2), 199-213.


\bibitem[Guttman (1998)]{Guttman.1998}
Guttman, N. B. (1998). \textit{Comparing the palmer drought index and the standardized precipitation index 1}. JAWRA Journal of the American Water Resources Association, \textbf{34}(1), 113-121.


\bibitem[Joe (2005)]{Joe.2005}
Joe, H. (2005). Asymptotic efficiency of the two-stage estimation method for copula-based models. Journal of Multivariate Analysis, 94(2), 401-419.


\bibitem[kim at al. (2007)]{kim.et.al.2007}
Kim, G., Silvapulle, M. J., and Silvapulle, P. (2007). \textit{Comparison of semiparametric and parametric methods for estimating copulas}. Computational Statistics and Data Analysis, \textbf{51}(6), 2836-2850.


\bibitem[Kullback and Leibler (1951)]{Kullback.and.Leibler.1951}
Kullback, S., and Leibler, R. A. (1951). \textit{On information and sufficiency}. The Annals of Mathematical Statistics, \textbf{22}(1), 79-86.


\bibitem[Ma et al. (2013)]{Ma.et.al.2013}
Ma, M., Song, S., Ren, L., Jiang, S., and Song, J. (2013). \textit{Multivariate drought characteristics using trivariate Gaussian and Student t copulas}. Hydrological processes, \textbf{27}(8), 1175-1190.


\bibitem[McKee et al. (1993)]{McKee.et.al.1993}
McKee, T. B., Doesken, N. J., and Kleist, J. (1993, January). \textit{The relationship of drought frequency and duration to time scales}. In Proceedings of the 8th Conference on Applied Climatology, \textbf{17}(22), 179-183.

%


\bibitem[Millar (1981)]{Millar.1981}
Millar, P. W. (1981). \textit{Robust estimation via minimum distance methods}. Zeitschrift für Wahrscheinlichkeitstheorie und verwandte Gebiete, \textbf{55}(1), 73-89.


\bibitem[Mishra and Singh (2010)]{Mishra.and.Singh.2010}
Mishra, A. K., and Singh, V. P. (2010). \textit{A review of drought concepts}. Journal of hydrology, \textbf{391}(1-2), 202-216.






\bibitem[Nagler (2018)]{Nagler.2018}
Nagler, T. (2018). \textit{kdecopula: An R Package for the Kernel Estimation of Bivariate Copula Densities}. Journal of Statistical Software \textbf{84}(7), 1-22.






\bibitem[Rao et al. (1975)]{Rao.et.al.1975}
Rao, P. V., Schuster, E. F., and Littell, R. C. (1975). \textit{Estimation of shift and center of symmetry based on Kolmogorov-Smirnov statistics}. The Annals of Statistics, 862-873.


\bibitem[Read and Cressie (2012)]{Read.Cressie.2012}
Read, T. R., and Cressie, N. A. (2012). \textit{Goodness-of-fit statistics for discrete multivariate data}. Springer Science and Business Media.


\bibitem[Shiau (2006)]{Shiau.2006}
Shiau, J. T. (2006). \textit{Fitting drought duration and severity with two-dimensional copulas}. Water resources management, \textbf{20}(5), 795-815.




\bibitem[Sklar (1959)]{Sklar.1959}
Sklar, M. (1959). \textit{Fonctions de repartition an dimensions et leurs marges}. Publ. inst. statist. univ. Paris, \textbf{8}, 229-231.

\bibitem[Song and Singh (2010a)]{Song.and.Singh.2010a}
Song, S., and Singh, V. P. (2010). \textit{Meta-elliptical copulas for drought frequency analysis of periodic hydrologic data}. Stochastic Environmental Research and Risk Assessment, \textbf{24}(3), 425-444.


\bibitem[Song and Singh (2010b)]{Song.and.Singh.2010b}
Song, S., and Singh, V. P. (2010). \textit{Frequency analysis of droughts using the Plackett copula and parameter estimation by genetic algorithm}. Stochastic Environmental Research and Risk Assessment, \textbf{24}(5), 783-805.


\bibitem[Tsukahara (2005)]{Tsukahara.2005}
Tsukahara, H. (2005). \textit{Semiparametric estimation in copula models}. Canadian Journal of Statistics, \textbf{33}(3), 357-375.


\bibitem[Wei\ss (2011)]{weib.2011}
Wei\ss, G. (2011). \textit{Copula parameter estimation by maximum-likelihood and minimum-distance estimators: a simulation study}. Computational Statistics, \textbf{26}(1), 31-54.


\bibitem[Wong et al. (2008)]{Wong.et.al.2008}
Wong, G., Lambert, M. F., and Metcalfe, A. V. (2007). \textit{Trivariate copulas for characterisation of droughts}. Anziam Journal, \textbf{49}, 306-323.


\end{thebibliography}


\end{document}